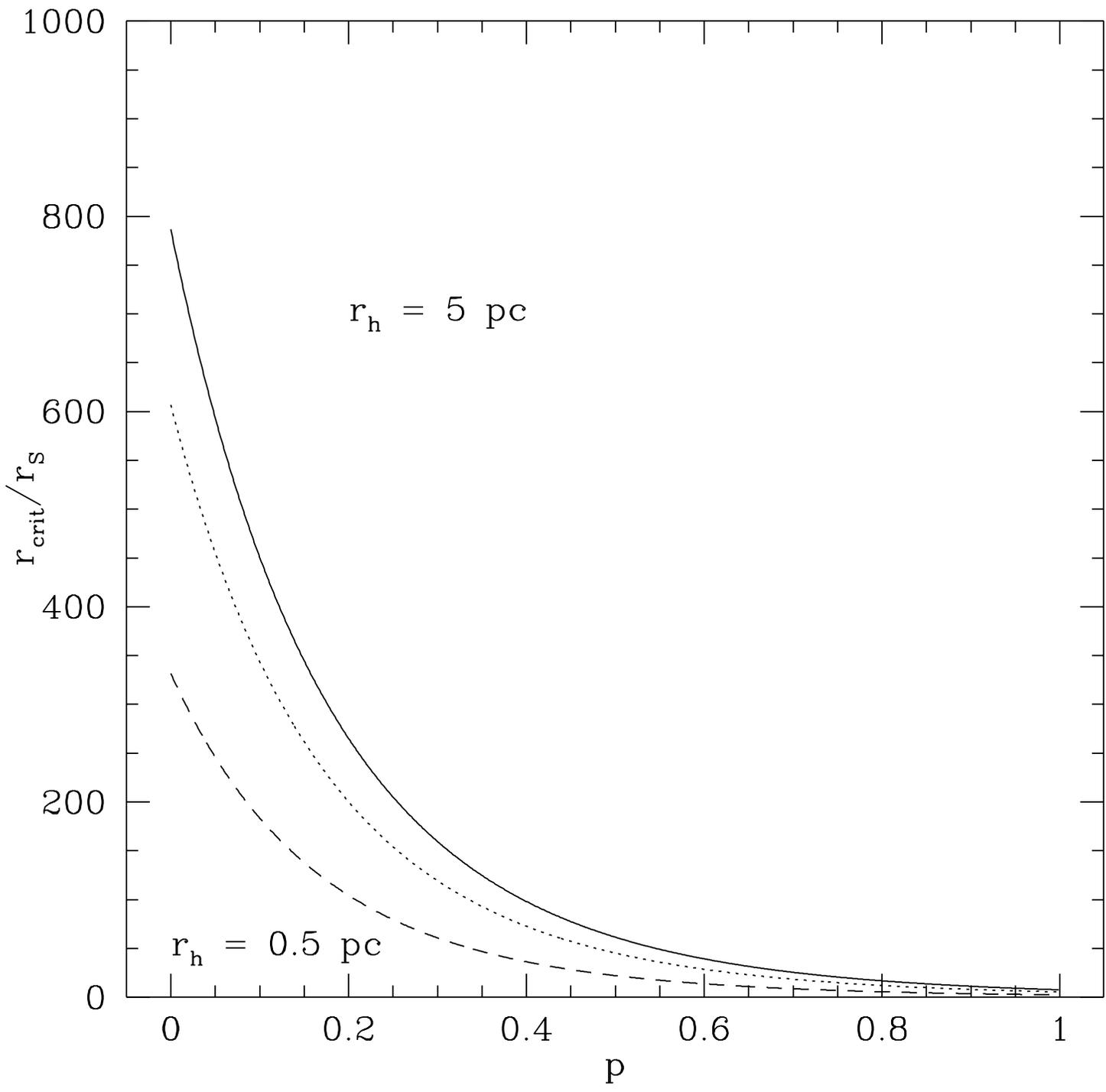

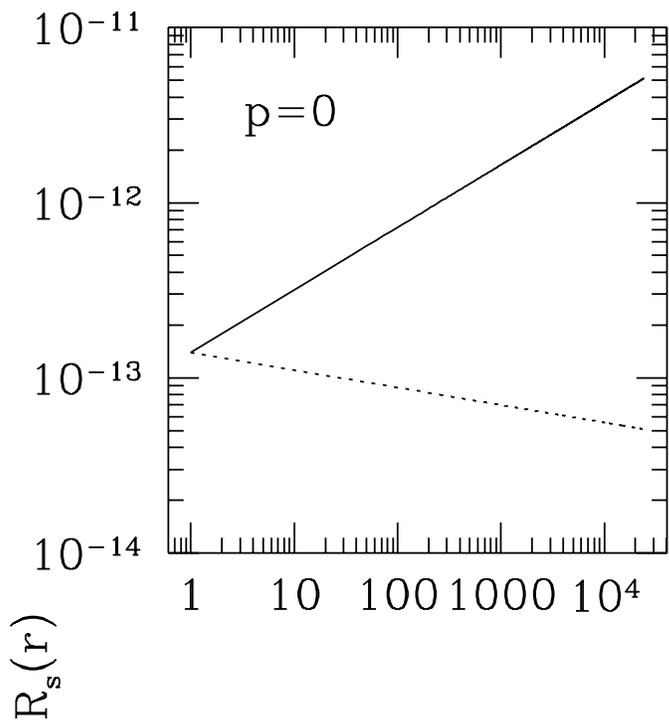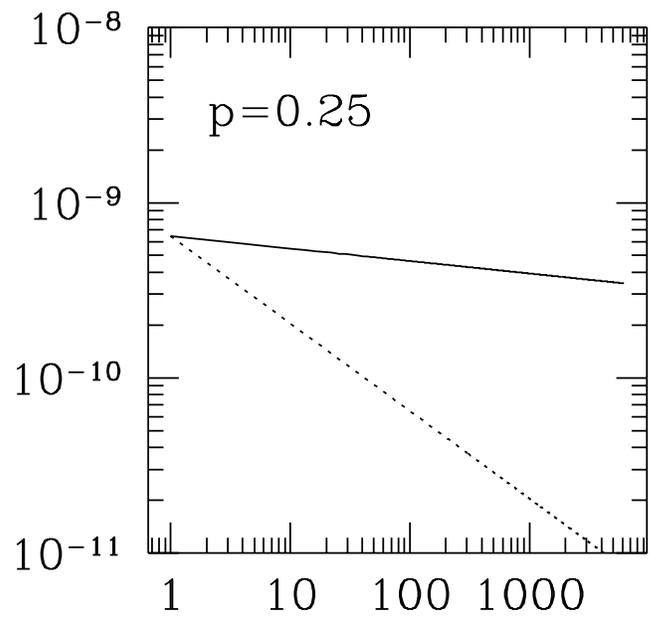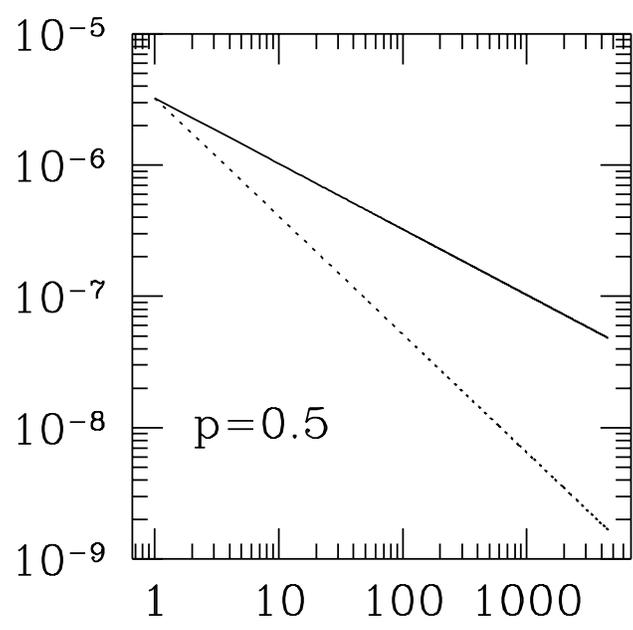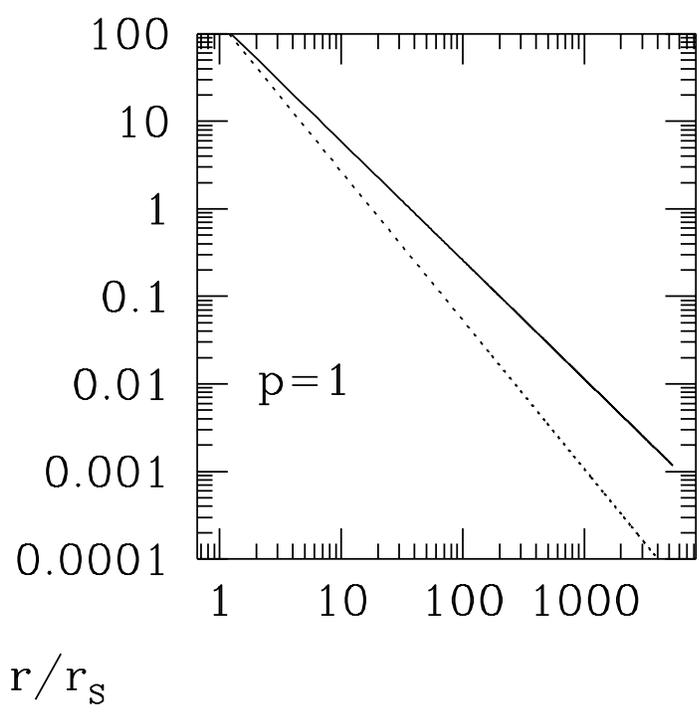

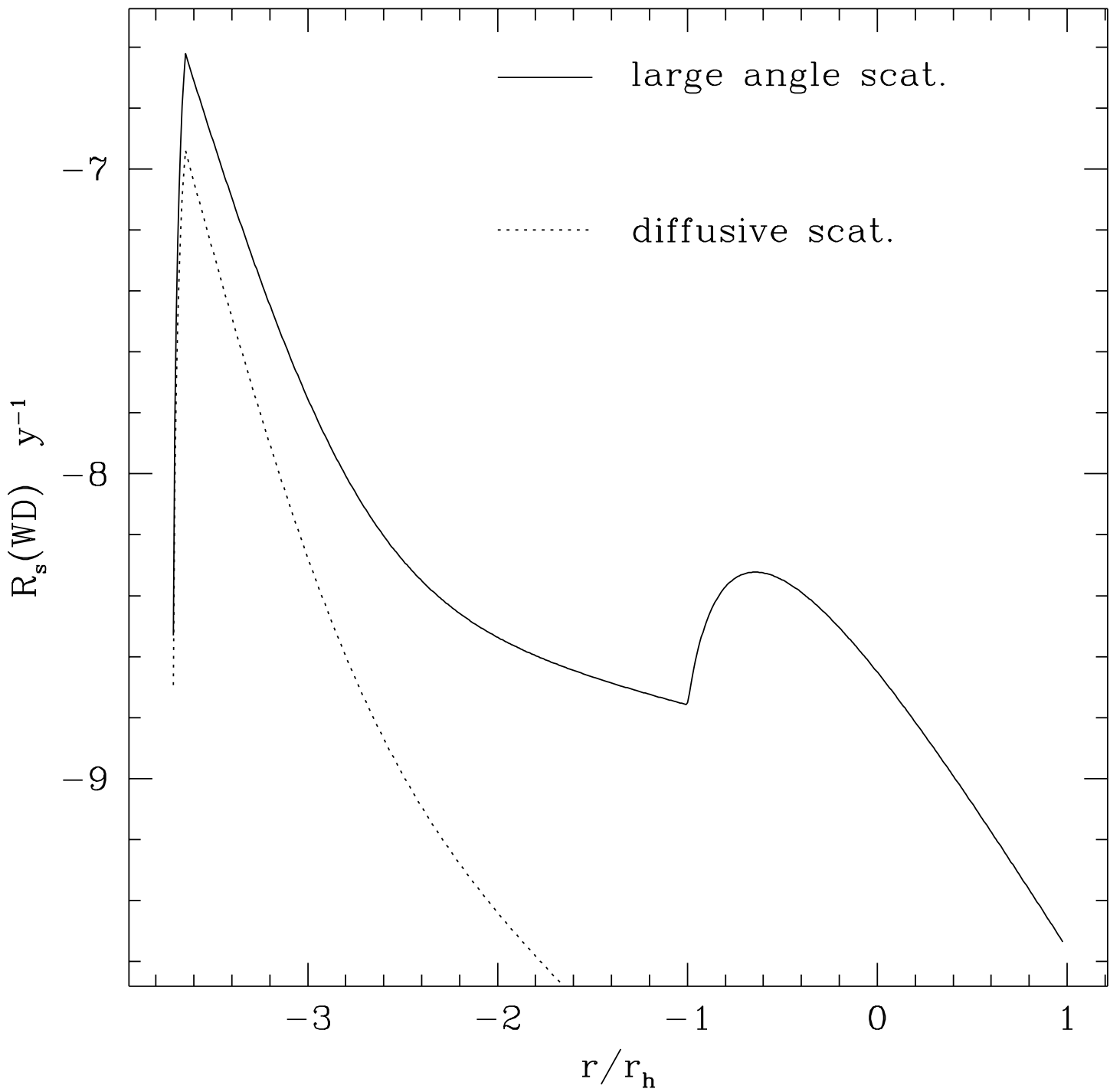

# Capture of stellar–mass compact objects by massive black holes in galactic cusps

S. Sigurdsson and M.J. Rees[1]
*Institute of Astronomy, Madingley Road, Cambridge CB3 0HA*
[1]*Institute for Advanced Study, Princeton, NJ 08540*

**ABSTRACT**

A significant fraction of the stellar population in the cusp around central black holes of galaxies consists of compact remnants of evolved stars, such as white dwarfs, neutron stars and stellar mass black holes. We estimate the rate of capture of compact objects by massive central black holes, assuming most spiral galaxies have a central black hole of modest mass ($\sim 10^6 \, \mathrm{M}_\odot$), and a cuspy spheroid. It is likely that the total capture rate is dominated by nucleated spirals. We estimate the flux of gravitational wave radiation from such coalescences, and the estimated detectable source count for proposed space–based gravitational wave observatories such as LISA. About one event per year should be detectable within $1 \, \mathrm{Gpc}$, given very conservative estimates of the black hole masses and central galactic densities. We expect $10^2$–$10^3$ detectable sources at lower frequencies ($10^{-4}$ Hz) "en route" to capture. If stellar mass black holes are ubiquitous, the signal may be dominated by stellar mass black holes coalescing with massive black holes. The rate of white dwarf–white dwarf mergers in the cores of nucleated spirals is estimated at $\sim 10^{-6}$ per year per galaxy.

## 1 INTRODUCTION

There is increasingly strong evidence that massive black holes are found in the centers of galaxies (see eg. Kormendy & Richstone 1995, Tremaine 1995, Rees 1990). Observational evidence, and theoretical considerations, indicate that the masses, $M_h$, range from under $10^6 \, \mathrm{M}_\odot$ to over $10^9 \, \mathrm{M}_\odot$. Black holes are thought to form during, or possibly before the formation of the host galaxy. Early infall of low angular momentum material, and angular momentum transport in the disk of gas surrounding the proto–black hole fuels the early stages of black hole growth, with accretion rates $\gtrsim 1 \, \mathrm{M}_\odot \, \mathrm{y}^{-1}$ inferred (see eg. Rees 1990). The lifetime of the resultant quasar is uncertain, it is possible that less than 1% of galaxies host quasars, with lifetimes $\gtrsim 10^9$ y, but more likely that most galaxies undergo shorter periods of activity (see eg. Haehnelt and Rees 1993).

If the QSO duty cycle is low, then most galaxies must have undergone moderately short periods of accretion onto a central black hole, in order to account for the total QSO numbers, and a typical galaxy has a $M_h \gtrsim 10^7 \, \mathrm{M}_\odot$ central black hole. If the QSO/AGN duty cycle is high, then remnants in active galaxies are mostly super–massive and as many as $\gtrsim 99\%$ of (non–active) galaxies will have moderate mass ($M_h \sim 10^6 \, \mathrm{M}_\odot$) black holes that did not undergo prolonged strong accretion episodes. It is interesting to note that for our own galaxy and the nearby M32 dwarf elliptical, the best estimates of the mass of any central black hole is $\lesssim 3 \times 10^6 \, \mathrm{M}_\odot$ (Bender et al 1996, van der Marel et al 1996,

Eckart and Genzel 1996). The true black hole mass function probably lies between the extremes mentioned, with a tail of very high mass black holes in a few galaxies, and some unknown distribution of masses in normal galaxies. A useful guess for the black hole mass function assumes the mass of a central black hole is some fraction of the mass of the stellar spheroid, $M_h = f_g L$, where $f_g \approx 10^{-2.5} \mathrm{M}_\odot/\mathrm{L}_\odot$, and $L$ is the galaxy luminosity (Tremaine 1995). We assume for now that $M_h \propto L$ and a Schecter (1976) luminosity function

$$\Phi(L) = \Phi_0 \left(\frac{L}{L_*}\right)^\alpha e^{-(L/L_*)} \frac{dL}{L_*}, \quad (1)$$

where $L_* = 1.8 \times 10^{10} \mathrm{L}_\odot$, $\Phi_0 = 0.008 \, \mathrm{Mpc}^{-3}$ and $\alpha = -1.07$ (Tremaine 1995, Efstathiou et al. 1988).

In general a dense cusp of stars (and possibly gas) forms around any central mass in a galaxy (Peebles 1972, Young 1980). Stars in these cusps will occasionally come close enough to the black hole to collide with each other or be swallowed by the black hole (see eg Frank and Rees 1976, Hills 1975). For black holes with masses near the low end of the range expected ($M_h \sim 10^{6-7} \mathrm{M}_\odot$) stars and (sub)giants are tidally disrupted if they come close to the black hole; more massive black holes may swallow stars whole. We expect $\sim 10\%$ of stars in an evolved population, such as is observed in elliptical galaxies and the bulges of spiral galaxies to be white dwarfs(WDs), another $\lesssim 1\%$ of stars may be neutron stars(NSs) or even low mass ($7 - 100 \, \mathrm{M}_\odot$) black holes(LMBHs), depending on the initial mass function and fate of evolved massive stars, which is uncertain. These evolved stars will also encounter the central black hole, but won't be disrupted, and may be promising sources for gravitational waves detectable by proposed detectors such as LISA (Danzmann et al. 1993, Hough et al. 1995).

## 1.1 Stellar encounters with central black holes

A star, radius $r_*$, mass $m_*$, would be tidally disrupted by a black hole mass $M_h$ within a radius $r_T \approx 2 \times (M_h/m_*)^{1/3} r_*$. (Rees 1988, Evans and Kochanek 1989, Frolov et al. 1994). For a main sequence star of solar mass, encountering a $10^6 M_6 \, M_\odot$ black hole, $r_T \approx 1.4 \times 10^{13} M_6^{1/3}$ cm($\approx 50 r_S M_6^{-2/3}$. In general the central region of a galaxy will contain both main sequence and evolved stars, with a range of masses and radii. We need to assume some global initial mass function for the stars, here taken to be the Salpeter mass function, $dN_*/dm_* = m_*^{-1-x_*}$, where $x_* = 1.35$. A typical galaxy will contain an evolved population, with a turnoff mass for the main sequence of $0.7 - 1.0 M_\odot$, plus possibly an additional younger population. The evolved population will typically consist of about 0.2% neutron stars of $1.4 M_\odot$, about 7% white dwarfs, with masses from $0.5$–$1.3 M_\odot$, and possibly about 0.03% few $M_\odot$ black holes. As we discuss later, however, dynamical effects may alter these proportions within the central cusp.

A degenerate, dark, compact remnant of the same mass as a main sequence star has a much smaller radius. For white dwarfs, $r_{WD} \approx 0.01 \, R_\odot$, for neutron stars, $r_{NS} \approx 10^6$ cm, much less than $r_*$. Consequently the white dwarf tidal radius is smaller than that for a main sequence star of the same mass by approximately $r_{WD}/r_*$. Since the Schwarzschild radius of a black hole, $r_S = 3 \times 10^{11} M_h/(10^6 M_\odot)$ cm, increases linearly with $M_h$, we see that for a black hole with $M_h \gtrsim 10^6 M_\odot$, $r_S > r_T$ for white dwarfs. For the values of $M_h$ considered here, neutron stars are not subject to tidal disruption; nor, obviously, are any low mass black holes, which may also encounter the central black hole. Compact stellar remnants can orbit around galactic black holes inside the tidal disruption radius for stars.

The orbital period for a solar type star around black hole at $r_T$ is $P_T = 10^4$ s, independent of $M_h$. The characteristic frequency, $f_c$, of gravitational radiation associated with the orbit of compact remnants about the black hole inside $r_T$ is then $10^{-4}$–$10^{-2}$ Hz. Gravitational radiation of this frequency is not readily detectable by Earth bound detectors but is well matched to the expected sensitivity of space based detectors (Thorne 1987, 1995, Danzmann et al. 1993, Hough et al. 1995, Haehnelt 1994). The sensitivity of these detectors cuts off below $\sim 10^{-3}$ Hz, and the gravitational radiation rate by a star in the innermost stable orbit around a massive black hole scales as $M_h^{-2}$. For this reason we are primarily interested in moderate mass central black holes, $0.5 \lesssim M_6 \lesssim 5$.

The corresponding characteristic amplitude of the gravitational waves, $h_c$, for a source at distance $d_{Gpc} = d/1 \, \text{Gpc}$, $M_h \gg m_*$, is

$$h_c = 3.7 \times 10^{-24} \frac{1}{d_{Gpc}} \left(\frac{m_*}{M_\odot}\right) M_6^{2/3} \left(\frac{10^4}{P}\right)^{2/3} g^{1/2}, \quad (2)$$

where $P$ is the orbital period in seconds and $g$ is a geometric factor of order unity (Thorne 1987). For $d \gtrsim 1 \, \text{Gpc}$ a correction for cosmological curvature is necessary. The exact waveforms for eccentric orbits and resultant detectability of the radiation are more complicated (see for example Cutler and Flannagan 1995, Ryan 1995, Shibata 1994, Junker and Schäfer 1992).

What, then is the likely rate of such events? Hils and Bender (1995) recently considered the rate of capture in M32–like systems with isothermal cusps due to the diffusion of compact stars into the black hole, the diffusion being dominated interaction with main sequence stars on more loosely bound orbits. This work extends the consideration to non–isothermal systems and includes the contribution of both diffusion and large–angle scattering to the capture rate, and the mutual interaction of compact stars in tightly bound orbits.

## 2 CUSPS AND SWALLOWING RATES

Consider a massive black hole in the center of some galaxy. In general, the black hole will be surrounded by some (evolved) stellar population with a range of masses corresponding to some initial mass function and an associated population of evolved remnants stars, white dwarfs, neutron stars and low mass black holes. The stars will have some underlying density profile, with the density $\rho(r)$ typically well approximated by a (broken) power law (Kormendy and Richstone 1995) with some "break radius", $r_b$. For the galaxies which are the best candidates for harbouring black holes of mass $M_h \sim M_6$, $r_b$ is observed to be a few parsecs. We assume for now that the stellar distribution is spherical and the velocity distribution isotropic.

We define a radius of influence of the black hole, $r_h = GM_h/\sigma_c^2$, where $\sigma_c$ is the one dimensional dispersion for $r \gtrsim r_h$. For most systems of interest the dispersion is approximately constant for radii larger than but comparable to $r_h$, and $r_h$ is well defined. For the systems we will be interested in, $M_h \sim M_6$ and $\sigma_c \sim 150 \, \text{km s}^{-1}$. Defining $\sigma_{166} = \sigma_c/166 \, \text{km s}^{-1}$, it is useful to scale

$$r_h = 1 \frac{M_6}{\sigma_{166}^2} \, \text{pc}. \quad (3)$$

We define a characteristic dynamical time scale, $t_{dyn} = r_h/\sigma_c$, the time scale for a star with characteristic 1–D velocity to cross $r_h$; for systems considered here $t_{dyn} = 6000(r_h/\text{pc})/\sigma_{166})$ years. Inside $r_h$ the density profile is modified by the presence of the black hole (Peebles 1972, Bahcall and Wolf 1976, 1977, Cohn and Kulsrud 1978, Young 1980, Shapiro 1985, Murphy et al. 1991, Quinlan et al. 1995).

A stellar population "relaxes" on some characteristic time scale, $t_R = \sqrt{2} \sigma_c^3/\pi G^2 m_* \rho \log(1/2N_*)$, where $N_*$ is the number of stars (see eg. Binney and Tremaine 1987). Scaling to a normalised density $\rho_6 = \rho/10^6 \, \text{pc}^{-3}$, we find $t_R = 10^{10} \sigma_{166}^3/\rho_6$ years. Note that $t_R \lesssim t_H (= 1.5 \times 10^{10}$ years) is possible for reasonable values of $\sigma_c$, $\rho$.

If the lifetime of the stellar population in the center is longer than few times $t_R$, then the stellar population is relaxed; in particular mass segregation occurs due to dynamical friction. The lifetime of the population of stars at the center of the galaxy is obviously $\lesssim t_H$, indeed it may be $\ll t_H$ if there has been recent star formation with associated mass loss, or if the black hole arrived or formed in the center of the galaxy recently.

## 2.1 Cusps

A self–consistent cusp of stars forms around a central black hole with some density profile $\rho(r < r_h) \propto r^{-C}$ (Peebles 1972, Bahcall and Wolf 1976, 1977, Young 1980, Shapiro 1985, Quinlan et al. 1995). In general $C$ is a function of $r$, with different processes modifying the local slope at different radii and densities. The black hole may also induce a small anisotropy in the velocity distribution (Goodman and Binney 1984, Quinlan et al. 1995). The cusp extends to some inner radius, $r_{in}$, where the star are effectively destroyed or swallowed.

It is useful to parametrise the cusp slope as $C(r) = 3/2 + p(r)$. With $t_R \propto \sigma(r)^3 \rho^{-1}$ inside $r_h$, $\sigma(r) \propto r^{-1/2}$ and hence $t_R \propto r^p$. A black hole growing adiabatically in a flat isothermal core, with $t_R \gg t_H$, induces a cusp with $p(r < r_h) = 0$ (Young 1980, Quinlan 1995). This has the interesting property that $t_R$ is independent of radius inside $r_h$. For physical cores $C \leq 2.5$ and hence $p \leq 1$ for plausible cusps whether relaxed or not (see Quinlan et al. 1995, Sigurdsson et al. 1995, for discussion). For a relaxed population of equal mass stars in the central region $(t_R(r_h) \ll \min\{t_H, t_{form}\})$, $p(r) = 1/4$ and the relaxation time decreases slowly with $r$, while for a relaxed cusp with a range of stellar masses, $\bar{p} \sim 0.3$, with steeper cusps for the more massive stars (Bahcall and Wolf 1976, 1977, Murphy et al. 1991).

Two further quantities are fundamental in determining the structure of the cusp around the black hole. The collision radius, $r_{coll} = 7 \times 10^{16} M_6$ cm for main sequence stars (smaller by $r_{WD}/r_*$ for white dwarfs, effectively zero for our purposes for neutron stars and stellar mass black holes), is the radius at which star–star encounters cannot lead to large angle elastic scattering as the characteristic encounter velocities exceed the surface escape velocities of the stars (Frank and Rees 1976). Inside $r_{coll}$ two body relaxation is by definition ineffective and $t_R$ becomes large for the relevant population. Note that $r_{coll} \gg r_T$ for both stars and white dwarfs for the black hole masses of interest here. Another quantity is the black hole wandering radius, $r_w$, which measures the root mean square displacement of the black hole from the galaxy's center of mass due to the discreteness of the potential of the galaxy. For a black hole in an isothermal stellar core $r_w \approx \sqrt{m_*/M_h} r_b$ (Bahcall and Wolf 1976). For other core profiles the wandering radius is less well defined (Quinlan private communication). Assuming the graininess due to the stellar mass distribution dominates $r_w$, then for the density profiles and black hole masses of interest here, $r_h \gg r_w \gg r_T$. Consequently, depletion of the main sequence stellar population due to tidal disruption is not limited by loss–cone diffusion into $r_T$ as the black hole wanders on dynamical time scales short compared to $t_R$. The short period random walk of the central black hole does not affect our rate estimates because the rate estimates are stochastic, averaged over time and the ensemble of black holes. The short time scale fluctuations that move the black hole away from a compact remnant star that would otherwise have entered the gravitational wave loss–cone, are balanced on average by central black holes random walking into the path of a compact remnant star that would otherwise not have entered the gravitational radiation loss–cone.

Another critical radius is set by the radius at which mergers of main sequence stars with each other become important, $r_m$. The rate for star–star mergers per star is simply

$$R_{cc} = \frac{\pi r_*^2 \rho(r) \sigma(r)}{m_*}, \qquad (5)$$

for $r_m \leq r_h$, assuming $\sigma \gtrsim v_{esc}$, where $v_{esc}$ is the escape velocity at the surface of the star (true for $r \ll r_h$). For white dwarf–white dwarf and white dwarf–neutron star merger gravitational focusing increases $R_{cc}$ by a factor $(1 + 2Gm_*/r_* \sigma(r)^2)$. In our units,

$$R_{cc} = 10^{-3} \rho_6 \sigma_{166} \left(\frac{r_*}{R_\odot}\right)^2 \left(\frac{M_\odot}{m_*}\right) \left(\frac{r_h}{r}\right)^{-3-p} \text{ per } t_H. \qquad (6)$$

By definition, depletion of stars through merger becomes important when $R_{cc} \sim t_H^{-1}$, provided $t_R \gtrsim t_H$, else depletion only becomes effective at $r_{coll}$. For solar mass main sequence stars, $r_m \sim 10^{-1} r_h \approx r_{coll}$, as expected. Inside $\min\{r_m, r_{coll}\}$ the cusp induced by the black hole is flattened by the depletion of main sequence stars due to mergers, with the main sequence density profile flattening to $\rho(r < r_m) \propto r^{-D}$, where $D \sim 0 - 1/2$ ($p = -1 - -3/2$). (Lightman and Shapiro 1977, Murphy et al. 1991, Rauch 1995).

While the main sequence stellar density inside $r_m$ is flat, this radius has no special significance for the compact remnants which would maintain a steep profile inside $r_m$ (see eg Murphy et al. 1991), with

$$\rho_{cr}(r < r_m) = f_{cr} \rho(r_m) \left(\frac{r_m}{r}\right)^{-C}, \qquad (7)$$

where $f_{cr}$ is the fractional density of white dwarfs (or neutron stars or stellar mass black holes) at $r_m$, corrected for mass segregation if applicable. Compact remnants stars also merge with main sequence stars, as well as each other, the rate for merger with main sequence stars is smaller than the star–star merger rate by a factor $f_{cr}$, and simulations of such high speed mergers suggest in most cases the compact star will remain after the merger (Shara and Regev 1986, Benz et al. 1989, Ruffert 1992). Inside $r_{coll}(MS)$ the two body relaxation time scale for white dwarfs increases sharply as the effective density of bodies available for large angle scattering decreases by $f_{cr}$.

## 2.2 Dynamics of compact stars in the cusp

### 2.2.1 *The influence of the central black hole*

Consider now the dynamics of the stars in the cusp. The stellar orbits are dominated by the gravity of the central black hole, but are perturbed by the mutual interactions of the stars. Stars which venture too close to the black hole are either tidally disrupted, in the case of main sequence stars, or swallowed whole, in the case of compact stellar remnants. In the latter case, the flux of compact remnants into the black hole is due to stars in the cusp scattered into orbits that either plunge them directly into the black hole, or such that the peribothron is small enough for gravitational radiation to shrink the orbit more rapidly than interactions with other objects either scatter the star away from the black hole again, or put it on an orbit that plunges straight into the black hole.

To first non–zero order, gravitational radiation leads to a decrease in orbital energy, $E$, and angular momentum, $L$, with

$$\frac{dE}{dt} = -\frac{32}{5}\frac{G^4}{c^5}\frac{M_h^3 m_*^2}{r_p^5}f'(e), \tag{8}$$

and

$$\frac{dL}{dt} = \frac{32}{5}\frac{G^4}{c^5}\frac{M_h^3 m_*^2}{r_p^5}g'(e), \tag{9}$$

where $r_p = a(1-e)$ is the peribothron of an orbit with semi–major axis $a$, eccentricity $e$, $f'(e) = (1-e^2)^{3/2}(1+73/24e^2+37/96e^4)/(1+e)^5$ and $g'(e) = (1+7/8e^2)(1-e)^3/(1+e)^2$. Note that $f'$ is less sensitive than $g'$ to the eccentricity for $e \sim 1$ while both $E$ and $L$ are equally sensitive to $r_p$.

Frank and Rees (1976) considered the diffusion of main sequence stars in cusps around central black holes in galaxies with isothermal cores. We follow their argument for the scattering of compact stellar remnants for the full range of adiabatic and relaxed stellar cusps around central black holes. We consider stars inside the cusp at radii $r_S \ll r \lesssim r_h$. In order for the stars to be swallowed by the central black hole, their orbits must be within some critical radius $r_c(r)$. It is useful to define a "loss–cone", $\theta(r) = \sqrt{2r_{min}/3r}$, where $r_{min}$ is the peribothron distance for the star at $r$. As the stars orbit about the black hole, the orbits are scattered by the inhomogenous potential they move in; it is useful to consider two regimes; where the scattering angle is small compared to $\theta$, which we refer to as "diffusion", and scattering where the scattering angle is large compared to $\theta$, which we refer to as "kicks". The scattering in the respective regimes can be thought of as being due to the "Poisson noise" in the potential due to the discrete number of stars for diffusion, and as elastic scattering off individual stars in the case of "kicks". In the absence of a central black hole, the scattering in and out of the loss–cone would be symmetric, with the flux into the loss–cone balanced by the flux out. In the presence of a black hole there is an additional source of orbital evolution: the secular decay of the low angular momentum orbits due to gravitational radiation. Hence there is a net loss of stars to the black hole.

The ratios of the "scattering time scale", $t_{scat}$ and the time scale for decay through gravitational radiation, $t_{GW}$, to the orbital time scale, $t_{orb}(r)$, can be written

$$\frac{t_{scat}}{t_{orb}} = \left(\frac{M_h}{m_*}\right)^2 \frac{1}{N_*(r)}. \tag{10}$$

where $N_*(r) \propto r^{3/2-p(r)}$ is the number of stars interior to $r$. The time scale for decay through gravitational radiation is set by the energy radiation rate, given by

$$t_{GW} = \frac{64}{5}\frac{G^3 M_h^2 m_*}{c^5 a^4}f(e), \tag{11}$$

where $a$ is the semi–major axis of the stars orbit about the black hole, and $f(e) = (1+73/24e^2+37/96e^4)(1+e)^{-7/2}(1-e)^{-7/2}$. For the highly radial orbits we're interested in, $e \approx 1$, given $\theta = \sqrt{2/3}\sqrt{(1-e)}/$ we find

$$\frac{t_{GW}}{t_{orb}} = \frac{24\sqrt{2}}{85\pi}\left(\frac{3}{2}\right)^{7/2}\frac{M_h}{m_*}\left(\frac{r}{r_S}\right)^{5/2}\theta^7. \tag{12}$$

$t_{scat}$ is the time scale for scattering out due to "kicks"; the corresponding diffusion time scale is shorter by a factor $r_{min}/r = 3\theta^2(r)/2$. In order for a star to be scattered to a small enough peribothron that it will be swallowed due to gravitational radiation, we require $t_{scat} \geq t_{GW}$, or that $\theta \leq \theta_{crit}$. Hence for "kicks", we find

$$\theta_{crit} = \sqrt{\frac{3}{2}}\left(\frac{85\pi}{24\sqrt{2}}\right)^{1/7}\left(\frac{M_h}{m_* N_*(r)}\right)^{1/7}\left(\frac{r}{r_S}\right)^{-5/14}, \tag{13}$$

and for diffusion

$$\theta_{crit} = \sqrt{\frac{3}{2}}\left(\frac{85\pi}{24\sqrt{2}}\right)^{1/5}\left(\frac{M_h}{m_* N_*(r)}\right)^{1/5}\left(\frac{r}{r_S}\right)^{-1/2}. \tag{14}$$

For radii such that $\theta_{crit}(r) \geq 1$ the power–law cusp of stars is no longer present as stars are swallowed by the central black hole.

The rate, $R_s$, at which the stars are swallowed, either by scattering straight into the black hole, or by gradual shrinkage of their orbits is given by

$$R_s = \frac{N_*(r)\theta_{crit}^2}{t_{scat}}, \tag{15}$$

and we can solve for $R_s$ for a choice of cusp parameters and sum over the galaxy and black hole mass function for an estimate of the total rate. Substituting for $\theta_{crit}$

$$R_s(\beta) = \frac{3}{2}C_\beta^{2/\beta}N_*(r)^{2-2/\beta}\left(\frac{m_*}{M_h}\right)^{2-2/\beta}\left(\frac{r}{r_S}\right)^{-5/\beta}\frac{1}{P}, \tag{16}$$

where $\beta = 5$ for stars diffusing in the loss–cone, and $\beta = 7$ for stars undergoing large angle scattering out of the loss–cone, and $C_\beta^{2/\beta} = 3/2(85\pi/24\sqrt{2})^{2/7} \approx 2.7$ for $\beta = 7$, and $C_\beta^{2/\beta} = 3/2(85\pi/24\sqrt{2})^{2/5} \approx 3.4$ for $\beta = 5$.

### 2.2.2 *Single power law cusps*

If the cusp density is a constant power law, the number of stars, $N_*(r)$, is given simply by

$$N_*(r) = 4\pi \int_0^r \rho(r')r'^2 dr'; \tag{17}$$

normalising to $\rho(r_h) = \rho_6/10^6 M_\odot \text{pc}^{-3}$ we find

$$N_*(r) = \frac{1.2 \times 10^7}{3/2-p}\frac{\rho_6}{m_*/M_\odot}\left(\frac{r_h}{1 \text{ pc}}\right)^3\left(\frac{r}{r_h}\right)^{3/2-p}. \tag{18}$$

Assume we have a single power law cusp, that $p$ is independent of $r$ for $r < r_b$. It is useful to define $r_{crit}$, where $\theta_{crit}(r_{crit}) = 1$, then

$$r_{crit} = C_\beta^{\frac{1}{4+p}}\left(\frac{M_h}{m_*}\right)^{\frac{1}{4+p}} \times \left(\frac{3/2-p}{1.2 \times 10^7 \rho_6}\right)^{\frac{1}{4+p}}\left(\frac{r_h}{r_S}\right)^{\frac{3-2p}{8+2p}}r_S. \tag{19}$$

A plot of $r_{crit}$ vs. $p$ is shown in figure 1, for $M_6, \rho_6 = 1$, and $r_h = 0.5, 1.0$ and $5.0$ pc respectively. The curves are shown for $\beta = 7$; $r_{crit}$ is 25% larger for $\beta = 5$.

**Table 1.** Scaling of scattering rates. Assuming $\rho(r_h) = 10^6 M_\odot \, \text{pc}^{-3}$, $M_h = M_6$ and $\sigma = 166 \, \text{km s}^{-1}$, $r_{coll}/r_S = 2000$, $R_s(r) \propto (r/r_S)^\gamma$, $\gamma = (3 - 4p)/2 - (8 - 2p)/\beta$.

| $r_{crit}$ | $p$ | $\gamma$ | $t_R(r_{crit})/\text{y}$ | $R_s(r_{crit}) \, \text{y}^{-1}$ |
|---|---|---|---|---|
| $\beta = 5$ | | | | |
| 607 | 0 | $-1/10$ | $10^{10}$ | $7.3 \times 10^{-14}$ |
| 154 | 1/4 | $-1/2$ | $6 \times 10^8$ | $5.2 \times 10^{-11}$ |
| 45 | 1/2 | $-9/10$ | $2 \times 10^7$ | $1.1 \times 10^{-7}$ |
| 5.4 | 1 | $-17/10$ | $5 \times 10^3$ | 7 |
| $\beta = 7$ | | | | |
| 607 | 0 | $+5/14$ | $10^{10}$ | $1.4 \times 10^{-12}$ |
| 192 | 5/24 | 0 | $10^9$ | $1.6 \times 10^{-10}$ |
| 154 | 1/4 | $-1/14$ | $6 \times 10^8$ | $4.5 \times 10^{-10}$ |
| 45 | 1/2 | $-1/2$ | $2 \times 10^7$ | $4.8 \times 10^{-7}$ |
| 15 | 0.75 | $-13/14$ | $4 \times 10^5$ | $1.5 \times 10^{-3}$ |
| 5.4 | 1 | $-19/14$ | $5 \times 10^3$ | 14 |

**Figure 1.** $r_{crit}$ vs $p$ for single power law cusps for different $r_h$.

Then $r_{crit} \approx 600 r_S$ for $p = 0$, and $r_{crit} \approx r_S$ for $p = 1$, much less than $r_m$. Large angle scattering becomes ineffective for white dwarfs at $r_{coll} \approx 2000 r_S > r_{crit}$, but neutron stars and stellar mass black holes may undergo large angle scattering to $r \sim r_S$.

We consider the rate of capture for $\beta = 5, 7$ separately, still assuming a single power law cusp composed purely of compact remnants. Substituting in equation 17, we can write

$$R_s(\beta, r) = F(p, M_h) \left(\frac{r}{r_S}\right)^\gamma \tag{20}$$

where

$$\gamma = \frac{3 - 4p}{2} - \frac{8 - 2p}{\beta}, \tag{21}$$

and

$$F(p, M_h) = 10^{-4} C_\beta^{2/\beta} A(p)^{2 - 2/\beta} M_6^{-3/2 + 2/\beta} \\ \times \left(\frac{r_S}{r_T}\right)^{-3/2} \left(\frac{r_S}{r_h}\right)^{(3 - 2p)(1 - 1/\beta)}, \tag{22}$$

and

$$A(p) = \frac{1.2 \times 10^7}{3/2 - p} \frac{\rho_6}{m_*/M_\odot} \left(\frac{r_h}{1 \, \text{pc}}\right)^3. \tag{23}$$

Table 1 shows the resultant $r_{crit}$ and capture rates for different slope cusps.

The diffusion rate decreases with $r$ for all $p$. The large angle scattering rate increases with $r$ for $p < 5/24$. However $r_{crit}$ is much smaller than $r_{coll}$ for white dwarfs for the steeper cusps while for flat cusps the diffusion rate is relatively insensitive to $r/r_{crit}$. Note the rates shown in the last column are not the true rates as $r_{crit} < r_{coll}$ in all cases and large angle scattering is ineffective at these radii for white dwarfs. Figure 2 shows the capture rates for single power law cusps composed purely of compact remnants, for some different $p$, neglecting $r_{coll}$.

The fraction of stars that plunges directly into the black hole without gradual inspiral through gravitational radiation is $R_s(\beta = 5)/(R_s(\beta = 5) + R_s(\beta = 7))$. For single power law cusps, this can be a small fraction; for more realistic cusps we expect 1/3–1/2 the stars to plunge rapidly into the black hole, with the remainder undergoing a more gradual inspiral with the orbit eccentricity decreasing. Most stars plunge from $r \sim r_m \gg r_{crit}$ and enter $r < r_{crit}$ with eccentricities of $\sim 0.999$.

**Figure 2.** Plot of $R_s(r)$ for different $p$, assuming single power law cusps composed entirely of compact remnants and ignoring existence of $r_{coll}$. The solid line is for $\beta = 7$ (large angle scattering) and the dotted line is for $\beta = 5$ (diffusion).

For detection of gravitational radiation, the time the compact objects spend at orbital periods $P \lesssim P_4 (= P/10^4 \, \text{s})$ is critical. The time to decay for highly eccentric orbits, using the quadrupole approximation, is well approximated by

$$\tau_{GW} = 10^5 (1 - e^2)^{7/2} P_4^{8/3} \, \text{y} \tag{24}$$

for $M_6 = 1$, $m_* = M_\odot$, note also $\tau_{GW} \propto M_h^{-2/3}, m_*^{-1}$ (see eg. Rajagopal and Romani, 1995). For $e = 0.99$ at $r = 2000 r_S$ the time to inspiral is $\sim 10^4$ years, for $e = 0.999$ and $r = 4000 r_S$ the time to inspiral is only 30 years and the orbit has no time to circularise substantially before reaching periods $\lesssim P_4$. For steep (large $p$) cusps, where the rate is dominated by stars near $r_{coll}$, diffusion will random walk $e$ to somewhat lower values for a significant fraction of the stars, as discussed above. The eccentricity, $e_f$, as the star approaches capture is smaller with, $(1 - e_f) \sim 4 \times (1 - e_i)$, where $e_i$ is the initial orbital eccentricity after scattering, and the system may spend times of order $10^3$ years at periods $\sim P_4$. As discussed in a recent paper by Rauch and Tremaine (1996), resonant effects may enhance the relaxation rate of angular momentum in Keplerian potentials; however, this does not appear to be effective for the orbital parameters of concern here, because the relativistic precession of the relevant orbits (with small peribothrons) is fast enough to destroy the resonances. Eccentricity evolution inside the loss–cone is therefore small.

Stellar mass black holes may contribute strongly to the total observable rate. They spiral in an order of magnitude faster than white dwarfs and neutron stars and are observable for correspondingly shorter times at these frequencies. On the other hand, $h_c$ is an order of magnitude larger for the low mass black holes than for white dwarfs, and the volume for detection is $\sim 10^3$ times larger. If stellar mass black holes form in population II in significant numbers, $(f_{BH} \gg 10^{-4})$, they will provide a strong characteristic signal for LISA (see also Polnarev & Rees 1994).

In practice cusps do not have simple single power law density profiles as we discussed above, nor are they composed solely from compact remnants. We now consider the expected capture rates in real cusps.

## 3   REAL GALAXIES

There are two classes of galaxies whose centres are good candidates for harbouring central black holes in the appropriate mass range, which may be capturing compact objects at an interesting rate: nucleated spiral bulges, such as that of our own galaxy; and the cores of compact dwarf ellipticals like M32. Both have low dispersion, steep high density central cusps, and probably contain central black holes of $\sim 10^6$ M$_\odot$. M32 was discussed in detail by Hils and Bender (1995), assuming diffusion dominated the capture rate and that the central cusp was unrelaxed and isothermal ($p = 0$) It is likely that the core of M32 is in fact relaxed, with $p = 0.25$, although observation at radii $\sim 0, 1r_h$ are consistent with $p = 0$. Systems like M32 are particularly promising sources because of their large central densities and low central dispersion. We infer a capture rate of $\approx 3 \times 10^{-8}$ per year, dominated by large angle scattering from $r \approx 5 \times 10^3 r_S$. A somewhat smaller diffusion/collision rate of $1.8 \times 10^{-8}$ was estimated by Hils and Bender (1995). Unfortunately the space density of dwarf galaxies like M32 appears to be very small, probably as low as $10^{-5}$ Mpc$^{-3}$ (Kormendy private communication, see also Gebhardt et al. 1996), the total capture rate out to 1 Gpc is thus only about $10^{-2}$ y$^{-1}$.

### 3.1   Bulges of spirals

The space density of nucleated spirals with the appropriate bulge mass is higher $\sim 10^{-2.5}$ Mpc$^{-3}$, but the dynamics of their central regions are more complicated. At $r_h$ typical density profiles correspond to $p \approx 0.5 - 0.8$. For the densities and dispersions seen in nuclei of spirals, the cusp of main sequence stars is flattened due to stellar mergers at $r_m \approx 0.1 r_h$ as discussed above. Inside $r_m$ the main sequence density profile is flat, $\rho_{MS}(r) = \rho_0 (r/r_m)^{-1/2}$, $\rho_0 = \rho_6 (r_h/r_m)^{3/2+p} \approx 10^8$M$_\odot$ pc$^{-3}$. The relaxation time at $r_m$, is $t_R(r_m) \sim 3 \times 10^9$ years. The total number of main sequence stars inside $r_m$ is thus $N_{MS} \approx 5 \times 10^5$. The white dwarf–main sequence merger rate is smaller by a factor $f_{cr}$, and it is likely at the encounter velocities seen in these cores that a white dwarf would emerge relatively unscathed from any such merger. Thus we expect the white dwarf (and neutron star and low mass black hole) density profile to remain steep.

Including gravitational focusing, the cross–section for WD–WD mergers is $R_{cc}(WD) = 6 \times 10^{-5} (r_h/r)^{-2-p} t_H^{-1}$. Solving for $r_m(WD)$, and requiring $R_{cc}(WD) = 5$ as white dwarfs interior to $r_m(MS)$ are replenished on $t_R(r_m)$, we find $r_m(WD) \approx 10^{-2} r_h < r_{coll}(MS)$. The relaxation time scale for white dwarfs increases around $r_{coll}(MS)$ as WD–MS scattering becomes ineffective for two body relaxation, but then decreases like $r^p$ inside $r_{coll}(MS)$, with $r_R(WD) \lesssim t_H$ at $r_m(WD)$. Consequently, relaxation maintains the white dwarf density profile at $p \sim 0.3$ down to $r_{coll}(WD) \sim 5 \times 10^{-3} r_h \sim 2 \times 10^3 r_S$. Inside $r_{coll}(WD)$ we expect the density profile of the white dwarf population to flatten out to $\rho_{WD}(r) \propto (r/r_m(WD))^{-1/2}$.

The number of white dwarfs interior to $r_m(MS)$ is approximately $N_{WD} = 10^4$. The density of neutron stars and low mass black hole does not flatten due to mergers, but will level off due to relaxation inside $r_m(WD)$ to $p = 1/4$ profile. Note that with the density profile flattened the relaxation time no longer decreases inside $r_{coll}(WD)$ Inside $r_{coll}(WD)$ we thus find $t_R \sim t_H$ again and a strong ($p \gtrsim 0.5$) neutron star and low mass black hole cusp may persist all the way to $r_{crit}$.

The density profile of the neutron star and low mass black hole population follows that of the main sequence stars to $r_m(MS)$, $\rho_{NS/BH} = f_{NS/BH} \rho_6 (r/r_h)^{(-3/2-p)}$. At $r_m(MS)$ the profile may flatten to $p = 0.25 - 0.3$ as relaxation becomes effective. The larger value is appropriate if the main sequence population is evolved; the lower value is appropriate for a younger main sequence population (as appears to be the case in our own galaxy). The relaxed cusp profile persists to $r_{coll}(WD)$ at which point the $t_R$ is large and the profile may steepen to $p \sim 0.5$ again.

We calculated numerical large angle scattering rates and diffusion rates for piecewise power law cusps, where we allowed for the changes in the density profile inside $r_h$ due to stellar mergers, changes in local relaxation time scales, and ineffectiveness of collisional relaxation inside $r_{coll}$. Figure 3 shows the capture rates inferred as a function of radius for some characteristic profiles expected in the centres of nucleated spirals. Typical inferred WD–MBH capture rate, through large angle scattering from $r \sim 2 r_{coll}(WD)$ is $10^{-7}$ per year, for nucleated cusps with structural parameters like the Milky Way. Large angle scattering dominates the capture rate, and the total rate is dominated by white dwarfs at radii $\sim r_m(WD) \sim 3 \times 10^3 r_S$. With $\sim 10^4$ WDs inside $r_m$ the white dwarf population is not significantly depleted by mergers on time scales of $t_H$, and can be replenished by relaxation or stellar evolution as implicitly assumed in the derivation.

**Figure 3.** The rate of capture of white dwarfs by a $10^6 \, M_\odot$ central black hole in a canonical nucleated spiral galaxy, as a function of radius. The rate peaks strongly at $r \sim 2r_{coll}(WD)$ at $\sim 10^{-7}$ per year.

If the space density of these bulges is 0.003 Mpc$^{-3}$, we expect $\sim 1$ captures per year within 1 Gpc. At any one time we may expect $10^2$ systems in the process of capture within 1 Gpc, or $0.1 - 1$ system within 100 Mpc. Such systems would be easily detectable by LISA at 100 Mpc and might be detected near $10^{-3}$ Hz out to 1 Gpc. The optimal signal is expected from $M_6 \approx 3$: for smaller $M_h$ the gravitational wave amplitude becomes too small, for larger $M_h$ the frequencies become too low even for orbits close to $r_S$. Low central dispersion leads to higher capture rates, favouring systems with intrinsic density cusps similar to Hernquist profiles, low central dispersion and $p = 0.8$ inside $r_h$. Whether such systems are prevalent in nature is an open question.

If the IMF is flat and the neutron star fraction is higher than the 0.2% expected from a Salpeter slope mass function, then neutron star captures may be competitive with the white dwarf capture rate due to the steep density profile expected for the neutron stars at all radii. If WD-WD mergers lead to accretion induced collapse and neutron star formation the expected capture rate may be dominated by neutron stars formed in the cusp. $R_s(NS) \gtrsim 10^{-7}$ are possible for $f_{NS} \sim 1\%$, in which case the neutron stars contribute significantly to the total GW rate. However, neutron stars are expected to be born with natal kick velocities greater than $\sigma_{166}$ and the central potential may not be deep enough for many neutron stars to remain in the center after formation, depressing the effective $f_{cr}$. Low mass black holes are detectable to larger distances, but presumably occur in much smaller numbers. LMBHs are, however, presumed not to receive natal kicks.

If there is a substantial population of stellar mass black holes, the capture rate in spirals nuclei may be as high as $10^{-5}$ per year per galaxy for 10 $M_\odot$ black holes, with a global number fraction of $2 \times 10^{-4}$ for the black holes (see figure 4). With less than $10^3$ black holes in the central regions, such a rate is clearly not sustainable, but would be possible following a nuclear star burst, for a period of $10^8$ years or so. There is some evidence that the Milky Way underwent a nuclear starburst in the last $10^9$ years, if this is typical of nucleated spirals, the population averaged rate for stellar mass black hole mergers within 1 Gpc might be as high as $10^{-6}$ per year per galaxy, in which case we might detect $\sim 10$ such system coalescing per year with LISA. The expected initial eccentricities of the low mass black holes are $\sim 0.999$, and inspiral is rapid.

### 3.1.1 Binaries, anisotropy and triaxiality

There are likely to be some stellar binaries in galactic bulges. However, inside $r_h$ the density and dispersion are large and only binaries with semi-major axis $\sim r_*$ are hard enough not to be broken up by encounters with other stars. Such binaries cannot lead to a larger capture rate as the semi-major axis is $\lesssim r_{crit}$ and encounters with the binaries are no more effective in scattering them into the loss-cone than comparable single star scatterings. A significant fraction, $f_b$, of white dwarf-white dwarf binaries with semi-major axis $\sim R_\odot$ would lead to an enhanced WD-WD merger rate, by a ratio of $a_{WD-WD}/R_{WD} \sim 100 f_b$.

The black hole also polarizes the stellar distribution inducing a finite tangential anisotropy at small radii (see Quinlan et al. 1995 for discussion). There is consequently some bias towards circular orbits in the vicinity of the black hole which will lower the estimated capture rate by $\sim 10\%$. As other processes can induce comparably mild radial anisotropies (in particular ejection of single stars from the inner cusp by low mass black holes bound transiently to the central black hole), this does not change our estimate of the capture rate.

In the inner cusp the stellar distribution is forced towards sphericity by the black hole potential (there might in general be some modest rotational flattening). On larger scales the spheroid is most likely triaxial, possibly strongly triaxial. Orbit diffusion in triaxial potentials may be a strong factor in replenishing the loss-cones of central massive black holes; this effect is irrelevant, as the black hole mass is low enough that its "wandering" ensures the loss-cone remains filled independent of the shape of the stellar distribution.

As discussed by, for instance, Syer, Clarke and Rees (1991) a main sequence star on an eccentric orbit with peribothron somewhat larger than $r_T$ can cumulatively lose energy by impact on an accretion disk, so that it ends up on a tightly bound circular orbit. This process depends on the geometric cross section of the star and would be correspondingly less effective for a compact object. The presence of a modest mass gas accretion disk around the central black hole will therefor not affect our present estimates of the capture rate for compact objects.

### 3.2 Stellar mergers

The rate of WD-WD mergers is high within the black hole induced cusp. The merger rate per white dwarf at $r_m(WD)$, $R_{cc}(WD) = 3 \times 10^{-10}$ per year, by definition. There are $10^4$ white dwarfs interior to $r_m(WD)$ given the assumed cusp parameters. The integrated merger rate inside $r_m(WD)$, is $N_c(WD) \sim 10^{-6}$ y$^{-1}$, enough to destroy every white dwarf in the cusp in a Hubble time. As the cusp is relaxed at $r_m(WD)$ this is not a concern and the white dwarf population can be replenished both through local stellar evolution and replenishment from outside $r_m(WD)$. The expected number of WD–WD mergers within 100 Mpc is 10 per year, with a NS-WD merger expected once every few years within the same volume, depending on $f_{NS}$. As noted above, a substantial fraction of WD–WD binaries could increase the WD–WD merger rate by an order of magnitude.

**Figure 4.** The rate of capture of LMBHs by a $10^6$ M$_\odot$ central black hole in a canonical nucleated spiral galaxy, as a function of radius. The rate peaks strongly at $r \sim 10^3 r_S$ at $\sim 10^{-5}$ per year.

If WD–WD mergers lead to accretion induced collapse, rather than total disruption of the white dwarfs, then WD–WD mergers may lead to a substantial neutron star population in the centres of galaxies, producing up to $10^4$ neutron stars inside $r_m$ in a Hubble time. This would on on average double the estimated coalescence rates of white dwarfs and neutron stars with the central black hole.

The white dwarf–main sequence merger rate is $\sim 10^{-5}$ per year, and we assume a compact remnant remains after the collision. If the white dwarf is temporarily bound to the main sequence star after merger, a common envelope phase may ensue, producing a large, luminous stellar object with a white dwarf core. If such objects last $10^7$ years, we would expect $\sim 10^2$ to be observed at any one time in the inner 0.1 pc of a typical nucleated bulge.

## 4  CONCLUSIONS

We estimate the capture rate of compact stellar remnants by massive black holes in the centers of power law bulges of nucleated spirals and compact dwarf ellipticals.

The capture rate in M32–like ellipticals is high, and likely to be dominated by large angle scattering by other stars in the tightly–bound cusp. The net rate for M32 like systems is likely to be of order $10^{-8}$ per year, but the total rate in the galactic neighbourhood is probably small due to the dearth of such systems.

The total rate is likely dominated by the power law bulges of "ordinary" spiral galaxies such as our own Milky Way. The total expected rate is somewhat lower than naive estimates due to the flattening of the white dwarf density profile through WD–WD mergers, with likely event rates of order $10^{-8}$ per year per galaxy. However, due to their relatively high space density such systems dominate the total observable rate out to 1 Gpc.

We conservatively estimate a minimum of 0.1 captures per year out to 1 Gpc, with perhaps $10^2$ systems observable at low frequencies in the early stages of capture at any one time. Both the burst and periodic signals should be detectable by proposed gravitational radiation observatories such as LISA, out to few hundred Mpc, with a characteristic signal from the high eccentricity orbits. If low mass black holes are present in significant numbers, in the centers of galaxies, then the signal from LMBHs captured onto the central black hole through large angle scattering may be an order of magnitude larger still, averaged over the local population of galaxies.

The WD–MS merger rate is estimated at $10^{-7}$ per year, with $O(10^2)$ luminous stellar objects descended from such mergers observable in the inner nucleus at any one time. The WD–WD merger rate is estimated at $10^{-6}$ per year per galaxy, with several per year expected in the local supercluster. Such events may be detectable through X–ray or UV flaring by current space based observatories. Chemical contamination of the bulge from such events may also be significant (Khokhlov and Novikov in preparation).

## ACKNOWLEDGEMENTS

MJR gratefully acknowledges the support of the Royal Society. SS thanks the PPARC and EU DGXII for support.

*Capture of compact stars by black holes* 9Quinlan G.D., Hernquist L., Sigurdsson S., 1995, ApJ, 440, 554
Peebles, P.J.E., ApJ, 1972, 178, 371
Rajagopal, M., Romani, R.W., 1995, ApJ, 446, 543
Rauch, K.P., 1995, ApJ, submitted
Rauch, K.P., Tremaine, S., 1996, New Astronomy, submitted (astro-ph 9603018)
Rees M.J., 1990, Science, 247, 17
Rees M.J., 1988, Nature, 333, 523
Ruffert, M., 1992, A&A, 265, 82
Ryan, F.D., 1995, Phys. Rev. D., 52, 3159
Schechter, P., 1976, ApJ, 203, 297
Shapiro, S.L, Marchant, A.B., 1978, ApJ, 255, 603
Shapiro S.L., 1985, in Goodman J., Hut P., eds, Proc. IAU Symp. 113. Reidel Dordrecht, p. 373
Shapiro, S.L., Teukolsky, S.A., Black Holes, White Dwarfs, and Neutron Stars, Wiley (New York)
Shara, M., Regev, O., 1986, ApJ, 306, 543
Shibata, M., 1994, Phys. Rev. D, 50, 6297
Sigurdsson, S., Hernquist L., Quinlan, Q.D., 1995, ApJ, 446, 75
Syer, D., Clarke, C.J., Rees, M.J., 1991, MNRAS, 250, 505
Thorne, K.S., 1987, in Hawkins S.W., Israel W., eds, Three Hundred Years of Gravitation, Cambridge University Press, p. 330
Thorne, K.S., 1995, in van Paradijs, J., van den Heuvel, E., Kuulkers, E., eds., Proc. IAU Symp. 165, Compact Stars in Binaries. Kluwer Academic, p. 153
Tremaine, S., 1995, in "Unsolved Problems in Astrophysics", ed. J. Bahcall and J.P. Ostriker, Princeton University Press (Princeton), in press
van der Marel, R.P., Quinlan, G.D., Sigurdsson, S., de Zeeuw, T., Hernquist, L., 1996, in preparation
Young, P.J., 1980, ApJ, 242, 1232
**Figure 1.** $r_{crit}$ vs $p$ for single power law cusps for different $r_h$.

**Figure 2.** Plot of $R_s(r)$ for different $p$, assuming single power law cusps composed entirely of compact remnants and ignoring existence of $r_{coll}$. The solid line is for $\beta = 7$ (large angle scattering) and the dotted line is for $\beta = 5$ (diffusion).

**Figure 3.** The rate of capture of white dwarfs by a $10^6$ M$_\odot$ central black hole in a canonical nucleated spiral galaxy, as a function of radius. The rate peaks strongly at $r \sim 2r_{coll}(WD)$ at $\sim 10^{-7}$ per year.

**Figure 4.** The rate of capture of LMBHs by a $10^6$ M$_\odot$ central black hole in a canonical nucleated spiral galaxy, as a function of radius. The rate peaks strongly at $r \sim 10^3 r_S$ at $\sim 10^{-5}$ per year.

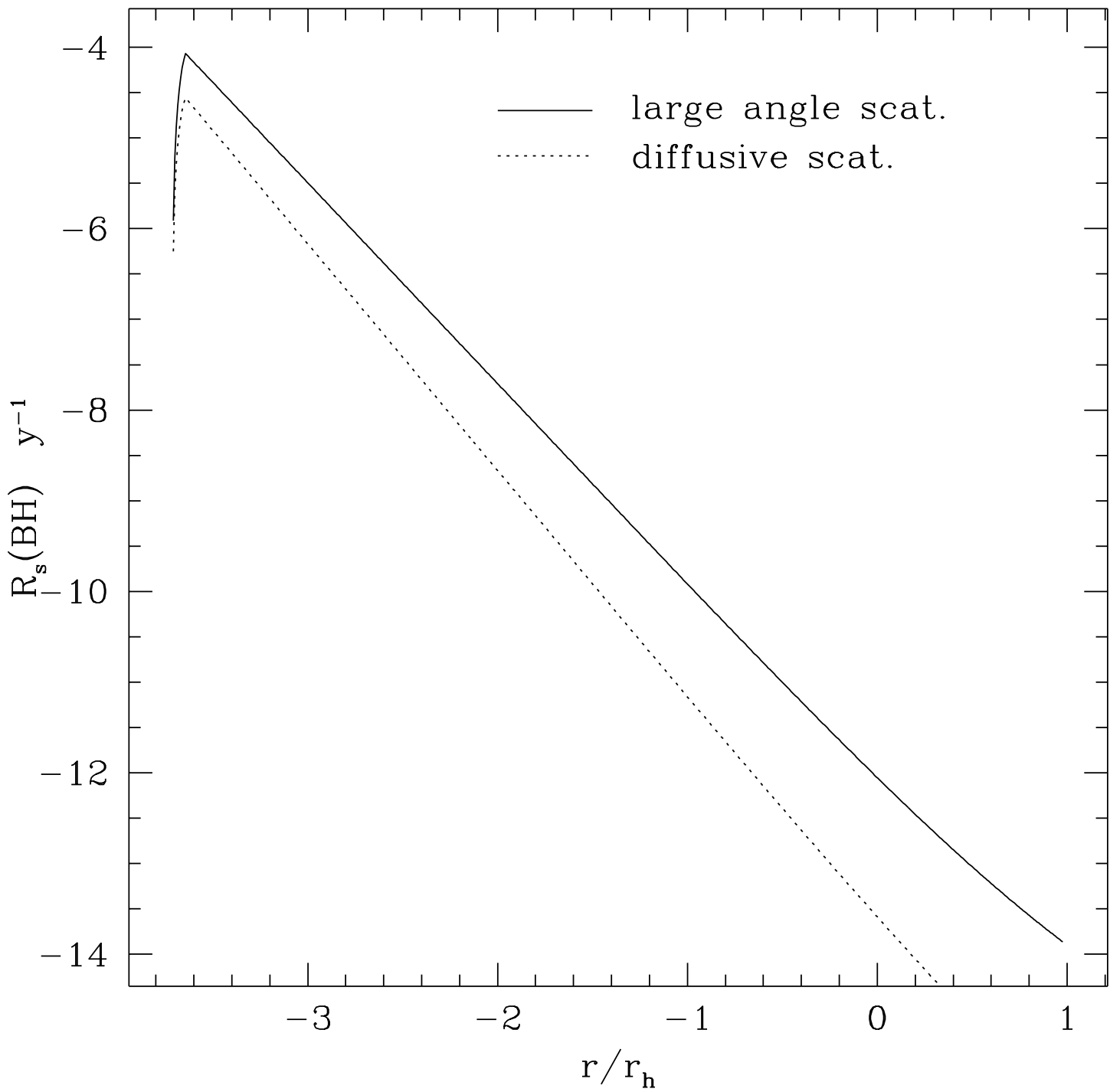